\documentclass[10pt,letterpaper,twocolumn]{article} 

\usepackage{ol2}
\usepackage{amsmath}

\usepackage{graphicx}
\usepackage{dcolumn}
\usepackage{bm}
\usepackage{amsxtra}
\usepackage[latin9]{inputenc}
\usepackage[small]{caption}

\usepackage{epstopdf} 
\usepackage[colorlinks=true, citecolor = blue, linkcolor = blue, pdfborder={0 0 0}]{hyperref}

\begin{document}

\twocolumn[ 
\title{Spectroscopy of a narrow-line optical pumping transition in dysprosium}
\author{M. Schmitt$^{1}$, E. A. L. Henn$^{1,2}$, J. Billy$^{1}$, H. Kadau$^{1}$, T. Maier$^{1}$, A. Griesmaier$^{1}$ and T. Pfau$^{1,*}$}
 \address{
 $^1$5. Physikalisches Institut, Universit\"{a}t Stuttgart, Pfaffenwaldring 57, 70569 Stuttgart, Germany \\
 $^2$Instituto de Física de São Carlos, Universidade de São Paulo, C. Postal 369, 13560-970, São 
Carlos, SP, Brazil  \\
 $^*$Corresponding author: t.pfau@physik.uni-stuttgart.de}
%
\begin{abstract}
We present measurements of the hyperfine coefficients and isotope shifts of the Dy I $683.731\,$nm transition, using saturated absorption spectroscopy on an atomic beam. A King Plot is drawn resulting in an updated value for the specific mass shift $\delta \nu_\mathrm{684,sms}^\mathrm{164-162}=-534\,\pm\,17\,\text{MHz}$. Using fluorescence spectroscopy we measure the excited state lifetime $\tau_{684}=1.68(5)\,\mu$s, yielding a linewidth of $\gamma_\mathrm{684} = 95\,\pm\,3\,\text{kHz}$. We give an upper limit to the branching ratio between the two decay channels from the excited state showing that this transition is useable for optical pumping into a dark state and demagnetization cooling.
\end{abstract}

\ocis{300.0300, 300.2530, 300.3700, 300.6460, 020.2930, 020.3260}
]


In atomic physics, rare-earth elements provide several applications: for instance dysprosium has been used to search for a possible variation of the fine-structure constant \cite{DyAlpha}. 
Transverse laser cooling of a beam of dysprosium atoms has been demonstrated \cite{DyOptPump} and recently quantum degenerate gases of bosonic and fermionic dysprosium \cite{DyBEC,Dyfermion} and bosonic erbium \cite{ErBEC} have been achieved, both with very large magnetic dipole moments $\mu_\text{Dy}=10 \, \mu_\text{B}$ and $\mu_\text{Er}=7 \, \mu_\text{B}$ with $\mu_\text{B}$ the Bohr magneton.
They join chromium \cite{Griesmaier,Beaufils,Lahaye.2007} in the quest to observe exotic many-body phenomena with strong dipolar interactions\cite{Lahaye.2009}.

One feature of strongly dipolar atoms is the possibility to implement demagnetization cooling \cite{Demagcooling}, which, in contrast to evaporative cooling \cite{DyBEC,ErBEC}, is lossless. Besides high dipolar relaxation rates \cite{DipolarRelaxation}, it requires a closed transition for optical pumping with a dark state in the energetic lowest magnetic substate.

In the case of dysprosium, one suitable candidate for such a process is the transition $4\mathrm{f}^{10}6\mathrm{s}^{2}$$ \, ^{5} \mathrm{I}_{8} \rightarrow 4\mathrm{f}^{9}(^{6}\mathrm{H}^{\mathrm{o}})5\mathrm{d}6\mathrm{s}^{2} \, ^{5} \mathrm{I}_{8}^\mathrm{o}$ at 683.731$\,$nm \cite{Martin}. There is not much known about this transition but the isotope shifts of the bosonic isotopes from which one can extract the Specific Mass Shift (SMS) for this transition \cite{japanese}.

In this paper we determine the fermionic hyperfine shifts as well as their isotope shifts for this transition. Additionally, we measure the lifetime of the excited state, from which we determine the transition linewidth. We give an upper limit to the branching ratio between the two possible decay channels and analyze the consequence of such branching ratio on the use of this transition for 'dark state pumping'.


To determine the isotope shifts and hyperfine structure of the transition, we performed saturated absorption spectroscopy on a beam of thermal dysprosium atoms. The atomic beam was obtained from a high-temperature effusion cell at 1250°C and propagated in an ultra high vacuum chamber ($10^{-9}\,$mbar). The atoms were observed through four viewports orthogonal to the atomic beam. One pair was used for the Doppler-free spectroscopy.

The 684$\,$nm light was produced by a commercial diode laser system (Toptica DL 100 / pro design) with tunable wavelength and a maximal output power of 17$\,$mW.  
The laser beam was split in two beams. One was sent to a confocal cavity (with a measured free spectral range of 150$\,$MHz) for frequency calibration. Nonlinearities in the frequency scan were smaller than 0.1\%. The second beam was used for spectroscopy and split into the pump and probe beams, which were then sent to the vacuum chamber in a quasi-counter-propagating configuration. The small angle between these beams allowed detecting the transmission of the probe beam.
In order to use lock-in detection of the transmitted signal, we chopped the pump beam by an acousto-optic modulator at $50\,$kHz.

\begin{figure}[t]
\centerline{\includegraphics[width=1\linewidth]{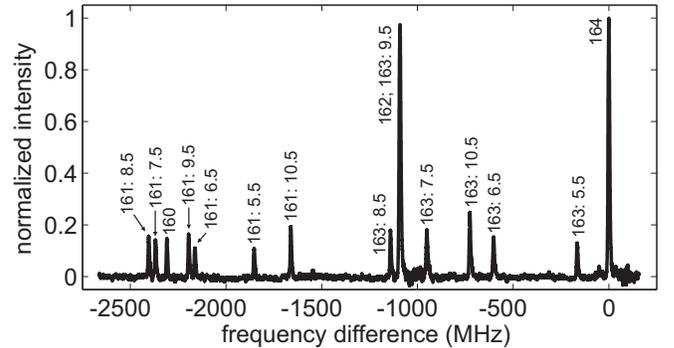}} 
\caption{Saturated absorption spectrum of the most abundant isotopes of dysprosium: the normalized intensity is plotted against the frequency difference to the highest peak ($^{164}$Dy). The spectrum is averaged over 100 single shots.}
\label{fig:spectrum}
\end{figure}

\begin{table}[b]
\caption{Hyperfine coefficients $A_e$ and $B_e$ of the excited state of the $684\,$nm transition for the fermionic Dy isotopes.}
\begin{minipage}{\linewidth}
\renewcommand{\footnoterule}{}
\renewcommand{\thefootnote}{\alph{footnote}}
\centering
{\begin{tabular*}{\linewidth}{@{\extracolsep{\fill}}ccc@{}} \hline
Isotope & $A_\text{e}\,$(MHz)\footnotemark[1] & $B_\text{e}\,$(MHz)\footnotemark[1]\\ \hline
163 & 152.56(2) & 2357(1) \\
161 & -108.84(2) & 2251(1) \\ \hline
\end{tabular*}}
\vspace{-10pt}
\footnotetext[1]{The uncertainties were evaluated as deviation of ($A_\text{e},B_\text{e}$) such that the hyperfine shift in Eq. \ref{eq:hfs} differs by half the FWHM of the fitted Lorentzian.}
\end{minipage}
\label{tab:coeff}
\end{table}

\begin{figure}[!t]
\centerline{\includegraphics[width=1\linewidth]{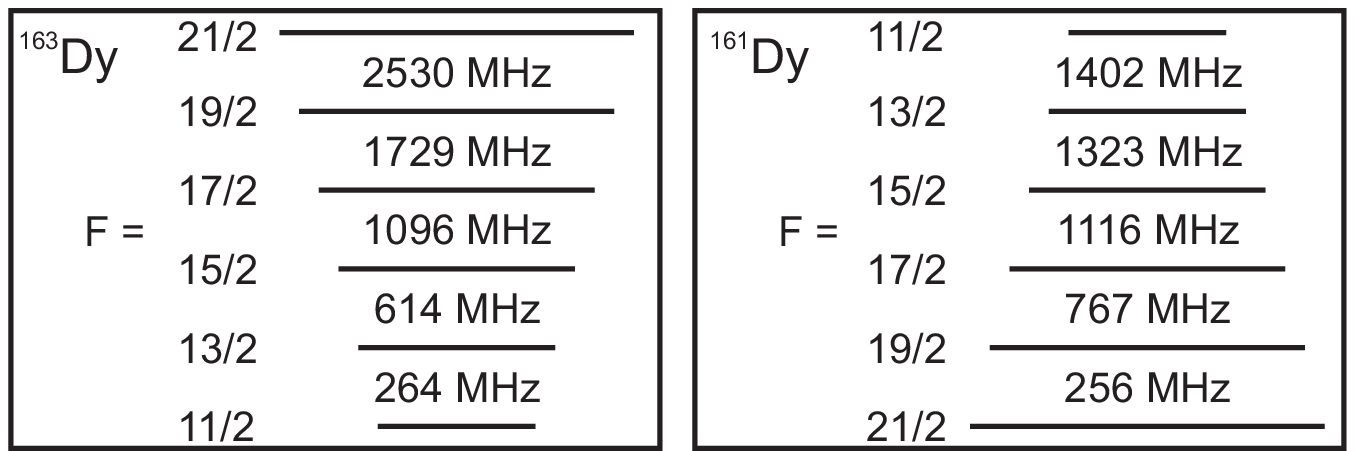}} 
\caption{Hyperfine levels of the excited state for the $^{163}$Dy and $^{161}$Dy fermionic isotopes.}
\label{fig:hfs684}
\end{figure}

Using this setup with a pump power $P_\mathrm{pump}=2.8\,$mW, a probe power $P_\mathrm{probe}=1.1\,$mW and $1/e^2$ beam radii $w_\mathrm{pump}\simeq2.9$ mm and $w_\mathrm{probe}\simeq1.6$ mm, we obtained the spectrum shown in Fig. \ref{fig:spectrum}.
The measured spectrum corresponds to a $J\rightarrow J$ transition with $J=8$. Thus the strongest transitions observed for the $^{161}$Dy and $^{163}$Dy fermionic isotopes correspond to $F\rightarrow F$ hyperfine transitions. 
The hyperfine splitting for a $F\rightarrow F$ transition is described by
\begin{flalign}
\nonumber
&\Delta \nu = (A_\text{e}\!-\!A_\text{g})\,(\mathrm{\textbf{I}}\cdot\mathrm{\textbf{J}})-\nu_\mathrm{iss} \\ 
&+(B_\text{e}\!-\!B_\text{g})\,\frac{3\,(\mathrm{\textbf{I}}\cdot\mathrm{\textbf{J}})^{2}+\frac{3}{2}\,(\mathrm{\textbf{I}}\cdot\mathrm{\textbf{J}})-I(I \!+\!1)J(J \! + \! 1)}{2I(2I \! - \! 1)J(2J \! - \!1)} ,
\label{eq:hfs}
\end{flalign}
where ($A_\text{e},B_\text{e}$) are the electric dipole and magnetic quadrupole coefficients of the excited state and ($A_\text{g},B_\text{g}$) the ones for the ground state \cite{Childs}. $\nu_\mathrm{iss}$ describes the isotope shift in the transition with respect to the $^{164}$Dy bosonic isotope. The positions of the experimentally detected lines were assigned by a non-linear least-squares multi-Lorentzian fit to the spectrum shown in Fig. \ref{fig:spectrum}. To evaluate the hyperfine coefficients we use Eq. \ref{eq:hfs} for each line position. We solved this overdetermined equation system with a linear least-squares fit. The best-fit ($A_\text{e},B_\text{e}$) are shown in Table \ref{tab:coeff} while the isotope shifts for all observed isotopes are given in Table \ref{tab:iss}. The hyperfine levels of the excited state are shown in Fig. \ref{fig:hfs684}.
Based on the obtained values we determine the ratios $A^{163}_\text{e}/A^{161}_\text{e}=-1.4017(4)$ and $B^{163}_\text{e}/B^{161}_\text{e}=1.0471(9)$, which are similar to the ground state ratios\cite{Childs}, but we cannot exclude a possible hyperfine anomaly \cite{Greenlees}.

\begin{table}[b]
\caption{Isotope shifts relative to the $^{164}$Dy isotope, for the 684$\,$nm transition and the reference transition.}
\begin{minipage}{\linewidth}
\renewcommand{\footnoterule}{}
\renewcommand{\thefootnote}{\alph{footnote}}
\centering
\label{tab:iss}
{\begin{tabular*}{\linewidth}{@{\extracolsep{\fill}}ccc@{}} \hline
Isotope & 684$\,$nm-shifts\,(MHz)\footnotemark[2] & 457$\,$nm-shifts\,(MHz) \\ \hline
164 & 0 & 0\\
163 & -823(4) & 660(3)\\
162 & -1091(4) & 971(2)\\
161 & -2099(4) & 1744(3)\\
160 & -2309(4) & 2020(3)\\
\hline
\end{tabular*}}
\vspace{-10pt}
\footnotetext[2]{The uncertainties are the half FWHM of the fitted Lorentzian.}
\end{minipage}
\end{table}

\begin{figure}[b]
\centerline{\includegraphics[width=0.97\linewidth]{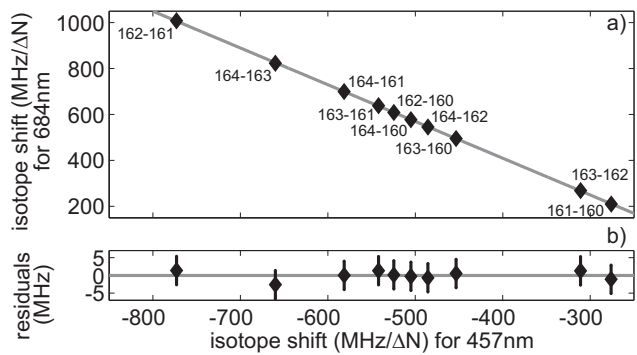}} 
\caption{King plot: (a) isotope shifts for the 684$\,$nm transition as function of the ones for the 457$\,$nm transition, both normalized to the atomic number difference $\Delta N$. The solid line corresponds to a linear fit to the data. (b) Fit residuals to the measured isotope shifts. The uncertainties correspond to half the FWHM of the fitted Lorentzian.}
\label{fig:kingplot}
\end{figure}

From the determined isotope shifts, we draw the corresponding King Plot \cite{King}, shown in Fig. \ref{fig:kingplot}, giving the isotope shifts of the considered transition as a function of the isotope shifts of a known reference transition, both normalized to the atomic number difference $\Delta N$.
We choose as reference the transition $4\text{f}^{10}6\text{s}^{2} \rightarrow 4\text{f}^{10}6\text{s}6\text{p}$ at 457$\,$nm \cite{Zaal}, which has a pure 'sp' excited state.

Using this King Plot analysis, we determine different contributions to the isotope shift: the field shift coming from the change in the nuclear charge distribution and the SMS resulting from the influence of correlations in the motion of the electrons \cite{Heilig}.
From a linear fit of the King plot, we extract its slope, equal to the ratio between electronic field shift factors $E_\mathrm{684}/E_\mathrm{457}=-1.601\pm0.003$, and the intersection $-231.1\pm1.8\,$MHz, which allows to calculate the SMS for the 684$\,$nm transition. For the two most abundant bosons, the SMS for the 457$\,$nm transition was measured to be $\delta \nu_\mathrm{457,sms}^\mathrm{164-162}=7\pm8\,$MHz \cite{Zaal} and we therefore find the corresponding SMS for the 684$\,$nm transition to be $\delta \nu_\mathrm{684,sms}^\mathrm{164-162}=-534\,\pm\,17\,\text{MHz}$.

Both the negative field shift factor ratio and the large negative value of the SMS for the 684$\,$nm transition show the different nature of this transition compared to the reference transition. Furthermore, the large negative SMS, indicating strong electrons correlations, is consistent with the typical values for $4\text{f}\rightarrow 5\text{d}$ transitions \cite{Zaal}. However, our measurement and the one of \cite{Zaal} slightly differ from the value given in \cite{japanese}, where a different reference transition was taken to draw the King plot.

The spectrum shown in Fig. \ref{fig:spectrum} does not allow us to extract the transition linewidth: indeed estimations of the transition saturation intensity and our beam intensities show that the lines are power-broadened.

We measured the linewidth of the 684$\,$nm transition by performing fluorescence spectroscopy with an oven temperature of 1350°C. A single laser beam ($P =13\,$mW and $1/e^2$ radius $w\simeq 1.9\,$mm), resonant with the frequency of the $^{164}$Dy isotope, is shone onto the atomic beam in a crossed configuration. We collected the emitted light with lenses and focussed it on a photodiode (Thorlabs - PDA36) with a verified bandwidth of $12.5$ MHz. By chopping the light with a period of $10\,\mu$s we observed the decay of the atoms back to the ground state.

\begin{figure}[t]
\centerline{\includegraphics[width=1\linewidth]{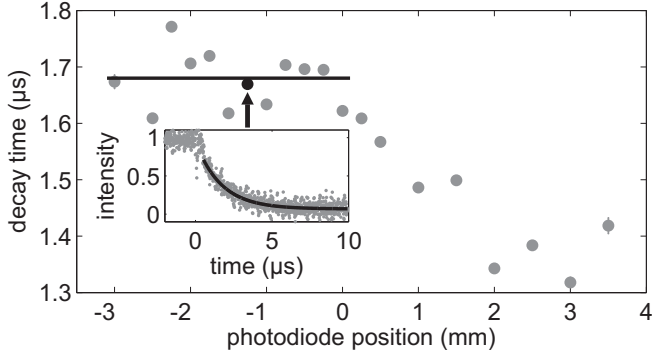}}
\caption{Measured decay time depending on photodiode position. The plateau at negative values indicates that the field of view is large enough not to be limited by transit time broadening. From a weighted mean value in this region we extract the lifetime of the excited state (solid line). The inset shows a fluorescence measurement with an exponential fit (solid line).}
\label{fig:lifetime}
\end{figure}

We averaged 10,000 single-shot measurements and applied a mean average of 100 points on the obtained decay curve to which we fit a single exponential decay, shown in Fig. \ref{fig:lifetime}.
We verified that we are not limited by transit time broadening arising from the movement of the atoms \cite{Ban.2005} out of field of view (FOV) before decaying back to the ground state. For this we moved the FOV by translating the photodiode relative to the lenses along the atomic beam direction and measured the effective decay time for different positions (see Fig. \ref{fig:lifetime}). We find a plateau over which the detected decay time is maximal and constant and verified with a CCD that the fluorescence distribution is 5 times narrower than the FOV. Thus transit time is not limiting the measurement in this region, where the measured decay constant is $\tau_{684}=1.68(5)\,\mu$s which yields a transition linewidth of $\gamma_\mathrm{684} = 95\,\pm\,3\,\text{kHz}$. 

There exists one additional dipole-allowed decay channel from the excited state to the metastable state $4\text{f}^{10}6\text{s}^2 \, ^{5}\mathrm{I}_{7}$, which was already observed in \cite{Conway} with a transition wavelength of $952.898\,$nm in air.
To measure the branching ratio between the two decay channels, we used a dichroic mirror, highly reflective at $684\,$nm and transmissive at $953\,$nm, and scanned the whole spectrum. With this measurement we are able to give an upper limit for the branching ratio of 1:100. This value agrees with the results obtained in \cite{Conway} comparing relative intensities. To determine if the $684\,$nm transition is suitable for dark state pumping we compare the branching ratio with the number of photons that are necessary to pump from $m_\mathrm{J}=8$ to $m_\mathrm{J}=-8$. Numerical calculations based on a recursive algorithm show that, on average, 16 photons are needed to pump an atom between the two extreme magnetic substates, which makes the $684\,$nm transition suitable for optical pumping into a dark state.


In this paper, we have investigated a narrow-line transition at 684$\,$nm in atomic dysprosium. We have determined the hyperfine structure and measured its linewidth to be $\gamma_\mathrm{684} = 95\pm3\,$kHz. By giving an upper limit of 1:100 to the branching ratio between the two decay channels from the excited state, we verify that this transition is suitable for dark state pumping and demagnetization cooling of atomic clouds of dysprosium. Similar schemes could be used with other atomic species with high magnetic moment: erbium using the transition at $622\,$nm ($\Gamma = 134$ kHz), thullium at $576\,$nm ($\Gamma = 207\,$kHz) and holmium at $661\,$nm ($\Gamma \approx 200\,$kHz).

This work was supported by the German Science Foundation (DFG) within SFB/TRR21. E.A.L.H. acknowledges support from the Alexander von Humboldt-Foundation and H.K. from Studienstiftung des deutschen Volkes.



\end{document}